# Utilization of H-reversal Trajectory of Solar Sail for Asteroid Deflection

Shengping Gong, Junfeng Li, Xiangyuan Zeng

*Tsinghua University, Beijing, 100084, CHINA*

**Abstract**: Near Earth Asteroids have a possibility of impacting with the Earth and always have a thread on the Earth. This paper proposes a way of changing the trajectory of the asteroid to avoid the impaction. Solar sail evolving in a H-reversal trajectory is utilized for asteroid deflection. Firstly, the dynamics of solar sail and the characteristics of the H-reversal trajectory are analyzed. Then, the attitude of the solar sail is optimized to guide the sail to impact with the object asteroid along a H-reversal trajectory. The impact velocity depends on two important parameters: the minimum solar distance along the trajectory and lightness number. A larger lightness number and a smaller solar distance lead to a higher impact velocity. Finally, the deflection capability of a solar sail impacting with the asteroid along the H-reversal is discussed. The results show that a 10 kg solar sail with a lead-time of one year can move Apophis out of a 600-m keyhole area in 2029 to eliminate the possibility of its resonant return in 2036.

## 1.Introduction

Earth is surrounded by Near Earth Asteroids(NEAs) and some are Potentially hazardous objects (PHOs), which are currently defined based on parameters that measure the object's potential to make threatening close approaches to the Earth. Large objects with an Earth minimum orbit intersection distance of 0.05 AU are considered PHOs. Objects with diameters of 5-10 m impact the Earth's atmosphere approximately once per year, with as much energy as the atomic bomb dropped on Hiroshima, approximately 15 kilotonnes of TNT. These ordinarily explode in the upper atmosphere, and most or all of the solids are vaporized. The rate of impacts of objects of at least 1 km in diameter is estimated as 2 per million years. Assuming that this rate will continue for the next billion years, there exist at least 2000 objects of diameter greater than 1 km that will eventually hit the Earth. Therefore, it is very necessary to prepare some new concepts for future use. There are usually two ways of deflecting dangerous NEAs. One way is to deflect the NEA using low thrust. The other way is to strike at the Asteroid at high relative velocity or a stand-off nuclear blast explosion. There are several ways of implementing low thrust deflection, such as propulsive devices in contact with the asteroid surface, surface ablation of the object using a laser or solar concentrator(Melosh, 1993; Gong, 2011), Yarkovsky effect(Joseph, 2002), exploitation of solar flux induced perturbations, mass driver, space tug and non-contact gravitational tractor(Lu &Love, 2005; Gong, 2009). Ahrens and Harris (1992) presented deflections methods by nuclear explosion radiation and surface nuclear explosion. Both utilize the energy released by the nuclear explosion to eject the mass of the asteroid that disturbs the velocity of the asteroid. McInnes(2004) considered deflecting the asteroid using a solar sail. A head-to-head impact is possible for a solar sail evolving in a retrograde orbit. The impact energy is comparable with that of the nuclear explosion for a relative velocity of impact larger than 60 km/s. Melsoh (1993) proposed a creative strategy that solar sail is used to focus sunlight onto the surface of the asteroid to generate thrust as the surface's layers vaporize.

For direct impact method, the required change in speed to be delivered to the asteroid in order to induce a change in position is a function of the time to impact (Izzo et al. 2005). From the results of Ahrens and Harris (1992), a velocity change of order 1 cm/s is required for a typical lead-time of order 10 years to deflect an asteroid for one Earth radius. The lead time is the time at which the impulse is applied prior to impact, and does not account for the time required to deliver the spacecraft to the asteroid. Given sufficient lead-time, it is possible for a relatively modest spacecraft to divert kilometer-sized asteroids. For example, to divert a 2 km asteroid with a 10 year lead-time requires an impact velocity of 10 km/s with a mass of order 60 tons. Raising the impact speed to 60 km/s leads to a significant reduction in the spacecraft mass to only 2.8 tons. To deliver the 2.8 ton spacecraft to a retrograde orbit at 1 AU from the Earth escape orbit requires a velocity increment of about 60 km/s. Using chemical propulsion with a specific impulse Isp of 450 s, about $2\times10^6$ tons of initial mass is required. A specific impulse Isp of 3000s still leads to a minimum initial mass of about 22 tons, neglecting trajectory gravity losses and the dry mass of the propulsion system. Solar sailing is a more attractive form of propulsion for such high-energy missions. A solar sail can deliver payloads into such high-energy retrograde orbits using the unique advantages of solar sailing. McInnes (2004) used a low performance solar sail of characteristic acceleration of 0.3 mm/s$^2$ to achieve a retrograde orbit. The solar sail spirals inwards from 1 AU to a close solar orbit of 0.25 AU. The 'orbit cranking' maneuver increases the solar sail orbit inclination in a monotonic fashion to obtain a retrograde orbit. The total transfer time to the retrograde orbit is about 10 years. Higher performance solar sails with characteristic accelerations of 0.5 mm/s$^2$ can achieve the mission in about 6.2 years. As the performance of the solar sail increases, the transfer time decreases.

In this paper, this new kind of retrograde impact trajectory is investigated. The H-reversal trajectory is achieved by reversing the momentum of the spacecraft using the solar radiation pressure. However, the orbit is not achieved by increasing the inclination. Instead, the momentum of the spacecraft is decreased continuously until the orbit is reversed. In this case, the spacecraft evolves in a retrograde hyperbolic orbit when impacting with the asteroid. Therefore, the impact velocity can be enhanced greatly. First, the solar sail dynamics is introduced and the retrograde orbit is achieved by reversing the momentum using solar radiation pressure force. Then, the asteroid deflection problem is stated and is converted into a parameter optimization problem. Finally, the deflection capability of a solar sail evolving in a H-reversal orbit is discussed.

## 2. Solar Sail Dynamics

An ideal plane solar sail is assumed. The lightness number of the sail is used to describe the solar radiation pressure acceleration that can be expressed as

$$\boldsymbol{f} = \beta \frac{\mu}{R^4} \left( \boldsymbol{R} \cdot \boldsymbol{n} \right)^2 \boldsymbol{n} . \quad (1)$$

where $\beta$ is the lightness number of the sail, $\boldsymbol{R}$ is the position vector of the sail relative to the Sun, $\mu$ is the solar gravitational constant, $\boldsymbol{n}$ is the unit vector directed normal to the sail surface. The performance of the sail is characterized by the sail lightness number, related to the density of the sail(McInnes, 2007) by

$$\beta = \frac{1.53}{\sigma}, \tag{2}$$

The unit for $\sigma$ is g/m$^2$.

A two body model is adopted and the gravitational perturbations of other celestial bodies are not included. Only the solar gravity and solar radiation pressure force(SRPF) exert on the solar sail. An inertial frame is used to discuss the dynamics of the solar sail. A system of nondimensional units is introduced for convenience. The distance unit is taken as astronomical unit, while the time unit is chosen such that the solar gravitational parameter is unitary. The transitions between the nondimensional units and international units are given in Table 1. With such a choice, the dynamical equation of motion in the inertial frame can be given by

$$\begin{cases} \dot{\boldsymbol{R}} = \boldsymbol{V} \\ \dot{\boldsymbol{V}} = -\frac{1}{R^3}\boldsymbol{R} + \beta\frac{1}{R^4}(\boldsymbol{R}\cdot\boldsymbol{n})^2\boldsymbol{n} \end{cases} \tag{3}$$

Table 1 Normalized units.

| Length (km) | Velocity (m/s) | Acceleration(m/s$^2$) | Time (day) |
|---|---|---|---|
| 1.496e8 | 29.24 | 8.5565e-5 | 58.1310 |

The sail acceleration vector can be described by two attitude angles, the cone angle $\alpha$ and clock angle $\delta$. Then, the solar radiation pressure acceleration can be written as

$$\begin{aligned} f_r &= \beta\cos^3\alpha / R^2 \\ f_t &= \beta\cos^2\alpha\sin\alpha\cos\delta / R^2 \\ f_h &= \beta\cos^2\alpha\sin\alpha\sin\delta / R^2 \end{aligned} \tag{4}$$

where $f_r$, $f_t$, $f_h$ are acceleration components along the radial, on-track and cross-track direction, as shown in Fig.1. Since the SRPF can not be sun-award, the cone angle belongs to $[-\pi/2, \pi/2]$ and the clock angle belongs to $[0, 2\pi)$. The transition matrix between the radial-on-track-cross track frame and inertial frame is determined by the position and velocity vector of the sail.

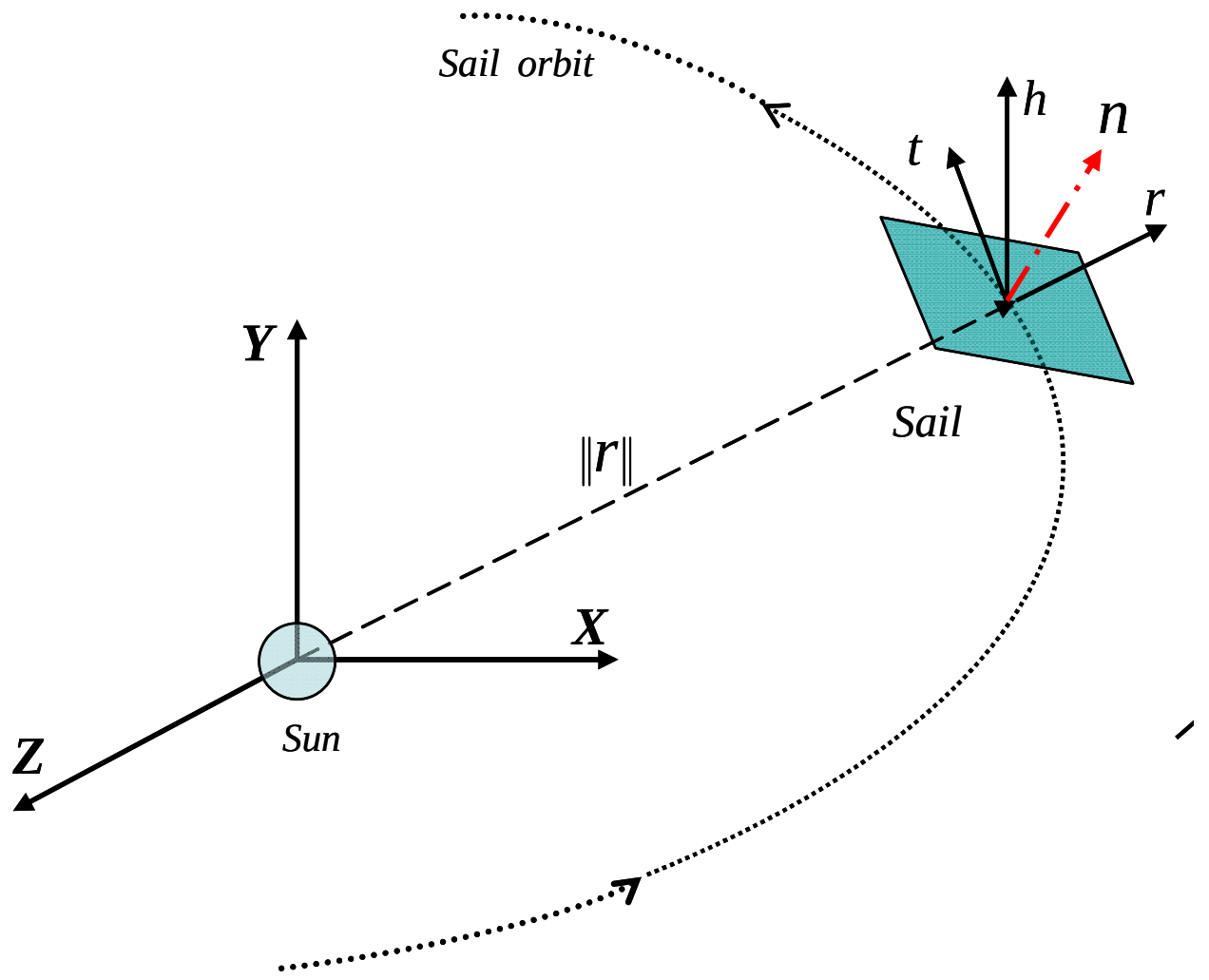

Fig.1 Orientation of the sail cone and clock angles

## 3. H-Reversal Trajectory by Solar Sail

Vulpetti(1996) was the first to address the H-reversal trajectory and he investigated both 2D and 3D H-reversal trajectories, including the dynamics and applications for interstellar missions (Vulpetti, 1996; 1997). Recently, Zeng (2011) discussed the applications of this trajectory. To achieve a high cruise speed for asteroid deflection, the sail has to gain enough energy to enter a hyperbolic orbit. A double or triple solar approach has been extended to pick up enough energy for low performance solar sail (McInnes, 2004). While for high performance solar sail, there have been two kinds of trajectory for solar approaches, that are the direct flyby and the H-reversal trajectory. For a H-reversal trajectory, the solar radiation pressure force is used to decrease the velocity of the solar sail and decrease the angular momentum at the same time until the velocity of the sail is parallel to the position vector at some point, where the angular momentum is zero. Further decrement of angular momentum will make the sail into a retrograde orbit. The sail decelerates further to a point where the velocity of the sail arrives at the minimum. Then, the sail begins to accelerate to the perihelion. Compared with direct multi-flyby cases, the transfer time can be greatly reduced. Mostly important, the escape velocity is higher than that of the flyby trajectory. Furthermore, the impact angle is always larger than 90 degrees since the reversal trajectory evolves in a retrograde orbit. A typical H-reversal is shown in Fig.2. The sail departs from the initial point A and the SRPF is used to decrease the velocity. Before arriving at the point C ( $h = 0$ ) the sail will get a maximum radius from the Sun at point B. Then the sail will pass through the perihelion D with negative angular momentum and escape from solar system to infinity.

For a given solar sail we will first identify the possibility of the fixed-cone-angle to produce the H-reversal trajectory. The fixed-cone-angle means the cone angle $\alpha$ is a constant during the whole trajectory. In order to seek the feasible region of $\alpha$ that generates H-reversal trajectory, a critical value of $\alpha$ corresponding to the trajectory with a perihelion of zero is identified. If the value of $\alpha$ is larger than the critical value, the H-reversal trajectory can be achieved. The range of $\alpha$ that can generate H-reversal trajectory increases with lightness number. Vulpetti's (1996) results show that the H-reversal trajectory only exists for high performance solar sail. He has given an approximate interval of $\lambda_r = \beta \cos\alpha$ within [0.5, 1) to realize the H-reversal mode trajectory. The H-reversal trajectory is also possible for a solar sail of lightness number less than 0.5 if the sail attitude angle is variable.

The sail in the same parameters in the H-reversal mode can pick up more energy than the direct flyby. This can be revealed from the fixed-Sun-angle trajectory. The work to change the mechanical energy of the sail is done by the SRPF. For H-reversal trajectory, the mechanical energy begins to increase before arriving at perihelion. With sharply increase of the mechanical energy, the sail is able to achieve a hyperbolic orbit when approaching the perihelion. However, for the direct flyby with only one solar approach the sail will always decrease its mechanical energy to achieve the required perihelion and then escape the solar system. This means the H-reversal trajectory begins to increase its mechanical energy before the perihelion point and can gain more energy than the direct flyby trajectory with the same solar sail.

Departing from the perihelion in a retrograde hyperbolic trajectory, the SRPF can be adjusted to guide a head-on impact with object asteroid. The whole transfer trajectory includes several phases. During the first phase, the pitch angle of the sail is adjusted to guarantee that the transverse component of solar radiation pressure force is used to decelerate the sail and reverse the angular momentum. The angular momentum of the final point of the first phase is zero. During the second phase, the solar radiation pressure force is used to guide the sail to the perihelion to gain kinetic energy. The third phase is from the perihelion to the impaction point. The trajectory control of this phase is difficult since the sail evolves in a hyperbolic trajectory and the velocity of the sail is very large. Therefore, the direction and opportunity of the hyperbolic trajectory at the perihelion should be optimized properly that the sail may impact with the object asteroid with a high relative velocity. In this paper, the pitch angle and clock angle during each phase are optimized to maximize the impact energy.

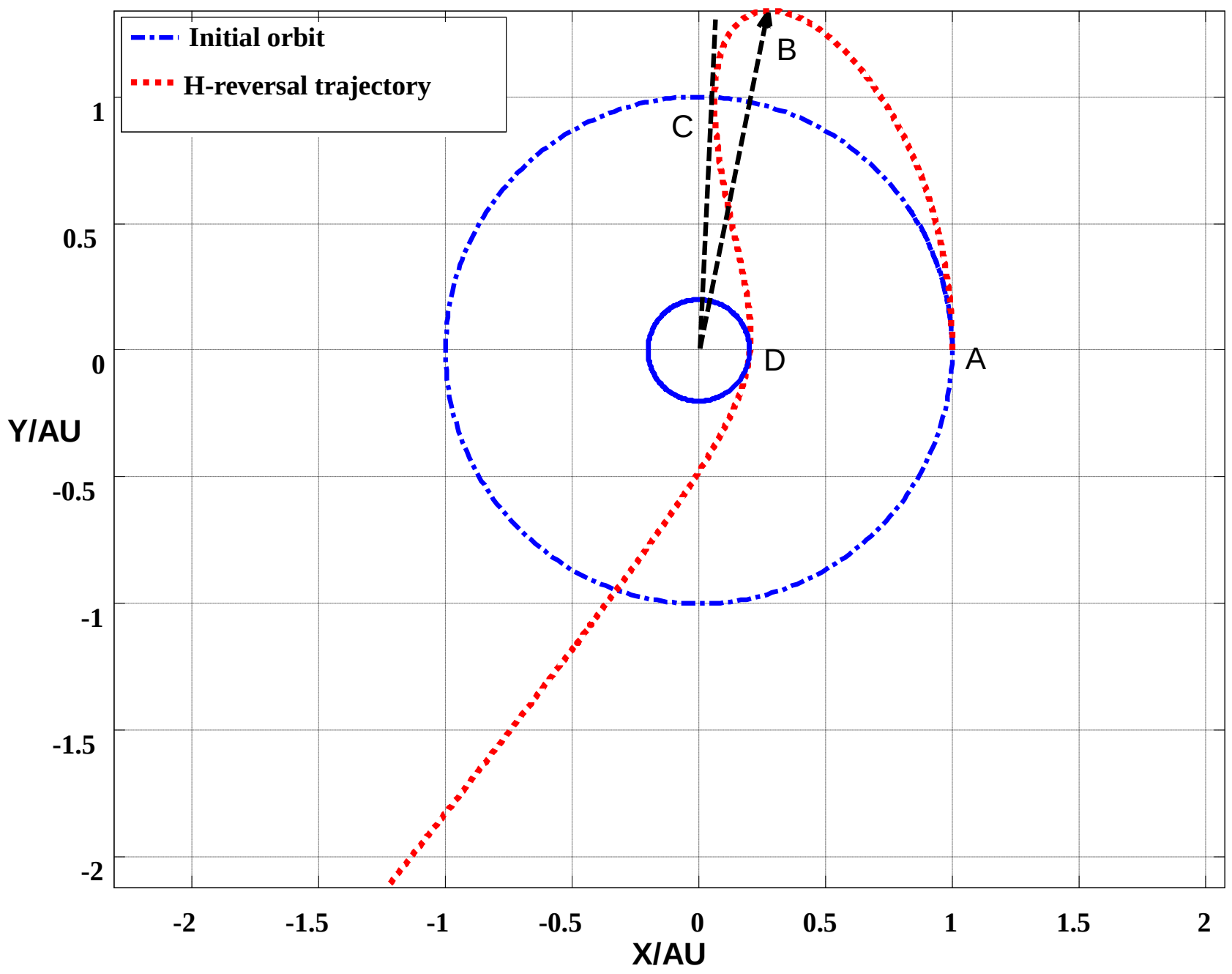

Fig.2 a typical H-reversal trajectory of solar sail

## 4. Asteroid Deflection Using Reversal Trajectory

### 4.1. Asteroid Apophis

Asteroid Apophis, known also by 2004 MN4, is a NEA with a size of 320 m and mass of about $4.6 \times 10^{10}$ kg. It was previously predicted that Apophis will pass about 36350 km above the Earth on April 13, 2029. Recent observations using Doppler radar at the giant Arecibo radio telescope in Puerto Rico confirmed that Apophis will swing by at about 32000 km above the Earth in 2029, but with a chance of resonant return in 2036.

The results in reference (Wie, 2007) show that a very small amount of velocity variation in 2026 is required to move Apophis out of a 600-m keyhole area in 2029 to eliminate the possibility of its resonant return in 2036. Keyholes are very small regions of the first encounter b-plane such that if an asteroid passes through them, it will have a resonant return impact with the Earth. In this paper, asteroid Apophis is used as an illustrative target asteroid assuming that it is going to pass through a 600-m keyhole in 2029. Both the Keplerian Elements of the Earth and Apophis in J2000 heliocentric ecliptic reference frame are used for simulations, as shown in Table 2.

Table 2 Classical Elements of the Earth and Apophis

| | Earth | Apophis |
|---|---|---|
| MJD | 5478 | 5478 |
| a (AU) | 1 | 0.92239 |
| e | 0.0167168 | 0.19104 |
| i (rad) | 0.000015454 | 0.05814 |
| Ω (rad) | 3.061420552 | 3.568535 |
| ω (rad) | 5.0198422625 | 2.205485 |
| M (rad) | 6.2347323914 | 0.361472 |

### 4.2 Optimization of the Deflection Trajectory

There are several parameters to be determined to achieve an impact trajectory from a H-reversal trajectory: the time of sail departing from the Earth, the time of the sail impacting with the asteroid and the attitude history of the sail. The attitude of sail determines the SRPF and is always treated as control variable. To maximize the impact velocity, an optimal control problem can be formed.

The literature on low-thrust and solar sail optimal control is extensive. Primarily, there are two methods for solving the resulting nonlinear optimal control problem: indirect methods and direct methods (Olympio, 2010). In an indirect method, first-order necessary conditions for optimality are derived from the optimal control problem via the calculus of variations. The primary advantages of indirect methods are their high accuracy and the assurance that the solution satisfies the first-order optimality conditions. However, indirect methods suffer from several disadvantages, including small radii of convergence, and the need for an accurate initial guess for the costate. A direct method is an alternate approach to identify the optimal transfer arc. In a direct method, the problem is parameterized by discretizing the trajectory and control variables, and explicit or implicit numerical integration schemes are used to satisfy the dynamical constraints. In this study, a direct shooting method is adopted to solve the optimal control problem. The control variables are parameterized along the transfer trajectory to maximize the impact velocity. The trajectory is divided into several segments. Over each segment, the control variables are treated as constant and the differential equations are integrated forward numerically. The control variables and the total fight time are optimized to ensure that the sail arrives at the target asteroid while maximizing the impact velocity. The whole trajectory design problem is converted into a parameters optimization problem.

As defined in Fig.1, the cone angle is between $-\pi/2$ and $\pi/2$ since the SRPF can not be sunward. Besides, the cone angle has an extra constraint to achieve a H-reversal trajectory before the sail arrives at the point of zero momentum. This constraint can also be satisfied by enforcing angular momentum. A small search space for the optimization parameters can reduce the optimization time and increases the probability of finding optimal solution. Therefore, the cone angle bound is enforced and the lower and upper bounds are determined by numerical methods. To avoid losing optimal solution, a looser bound is provided and the momentum enforcement guarantees the momentum reversal. The whole trajectory is divided into two sub legs by the zero momentum point, where the first leg is from the Earth to the zero momentum point that is discretized equally into $N_1$ segments and the second leg is from the zero momentum point to the impact point that is discretized equally into $N_2$ segments. During each segment, the cone angle and clock angle keep fixed. Now, the trajectory design problem is transformed into a parameter optimization problem, the optimization problem has $2(N_1+N_2)+2$ parameters, including the departure time from the Earth and arrival time at the asteroid. The optimization parameters can be given by

$$\boldsymbol{P}=\begin{bmatrix} t_0 & t_f & \alpha_1^1 & \delta_1^1 & \cdots & \alpha_{N_1}^1 & \delta_{N_1}^1 & \alpha_1^2 & \delta_1^2 & \cdots & \alpha_{N_2}^2 & \delta_{N_2}^2 \end{bmatrix}^T \quad (5)$$

where $t_0$ is the departure time from the Earth; $t_f$ is the arrival time at the asteroid; $\alpha_i^1, \delta_i^1$ ( i=1..$N_1$) are the cone angle and clock angle of the i$^{th}$ segment during the first leg, respectively; $\alpha_i^2, \delta_i^2$ ( i=1..$N_2$) are the cone angle and clock angle of the i$^{th}$ segment during the second leg, respectively. The bounds of optimization parameters can be specified. For example, the departure time $t_0$ can be assumed between 2015 and 2020. The arrival time is obtained by assuming that the total flight time is less than 4 years. As regards to the attitude angles, the cone angle of the first leg, that is $\alpha_i^1$, is between $\alpha_{\min}$ and $\alpha_{\max}$, where $\alpha_{\min}$ and $\alpha_{\max}$ are determined by numerical methods. The bounds of other attitude angles are determined by the constraint that the SRPF can not be sunward, that is $-\pi/2 \le \alpha_i^2 \le \pi/2$, $-\pi \le \delta_i^1 \le \pi$, $-\pi \le \delta_i^2 \le \pi$.

The H-reversal impact trajectory uses one solar approach to gain energy for high impact velocity. A smaller solar approach radius (the smallest distance from the Sun during the solar approach) leads to higher energy increment. Therefore, the impact velocity increases as approach radius decreases, which means that the maximum impact velocity is obtained when the sail tends to the Sun. However, a zero approach radius leads to the singularity of the dynamical equation and is also impossible for engineering practice. Therefore, a constraint on the approach radius should be added to the optimization problem to avoid the sail tending to the Sun. This process constraint can be given by

$$R(t) \ge R_{\min}, \quad t_0 \le t \le t_f \quad (6)$$

where $R(t)$ is the solar distance at time $t$ and $R_{\min}$ is the allowed minimum distance from the Sun during the total mission time.

The position error between the sail and the asteroid should be small enough to guarantee that the sail will impact with the asteroid. Usually, this condition is satisfied using an equality constraint.

$$\boldsymbol{R}\left(t_f\right)-\boldsymbol{R}_a\left(t_f\right)=0 \tag{7}$$

where $\boldsymbol{R}\left(t_f\right)$ and $\boldsymbol{R}_a\left(t_f\right)$ are position vectors of the sail at final time of the sail and asteroid, respectively.

The object function of the problem is to maximize the impact velocity that is the relative velocity between the sail and asteroid.

$$J=\left|\dot{\boldsymbol{R}}\left(t_f\right)-\dot{\boldsymbol{R}}_a\left(t_f\right)\right| \tag{8}$$

Now, the optimization problem can be stated as: find the optimal parameter vector $\boldsymbol{P}$ that maximizes the object function $J$ subjecting to inequality constraint given by Eq.(6) and equality constraint given by Eq(7).

The inequality constraint is a process constraint that is difficult to deal with direct method. In fact, the inequality constraint is equivalent to an equation constraint since the actual approach solar distance is always equal to the allowed minimum distance for this object function. The treatment of inequality and equation constraints is similar. The distance from the Sun is calculated for discrete points of each segment, which will not increase the computation burden since all the orbital parameters are calculated during integration of the dynamical equation. The object function is assigned a very small value if the constraint is violated.

The number of optimization parameters increases with the number of segments. Usually, a large number of segments leads to a difficulty of convergence to the optimal solution and a small number can not generate good results. In addition, large number means more sail attitude maneuvers. Therefore, the number of segments should be chosen properly to make the result close to true optimal solution but not too large for operation. To avoid local optimal results, the particle swarm optimization (PSO) (Kennedy & Eberhart, 1995) is employed to obtain the solution for a given number of segments. Through simulations of different segments, $N_1$=3 and $N_2$=6 is chosen since larger segments can not generate better solutions and smaller segments generate worse solutions. For PSO parameters, population size is 100 and maximum generation is 1000.

### 4.3 Simulations

The change in position is determined by the lead time and change in velocity. The lead time can be increased by reducing the transfer time while the change in velocity increases with the impact velocity. Therefore, the impact velocity and transfer time are the two most concerned parameters for the design problem. The lightness number of the solar sail describes the acceleration ability of the solar sail. A solar sail of large lightness number can gain large impact velocity by qualitative analysis. Another very important parameter is the minimum distance from the Sun during the solar approach. It is known that a smaller distance solar approach can gain larger energy. However, the minimum distance is limited by sail material limit bearing the hostile environment, such high temperature and all kinds of radiation. Therefore, a proper minimum distance can be chosen for the given sail material. To investigate how the lightness number and minimum distance influences the impact velocity and transfer time, the following cases are simulated. Firstly, solar sail of different lightness number are used to achieve the trajectory for a given minimum solar distance. Then, fix the lightness number and optimize the trajectory for different minimum solar distance constraints.

Figure 3 gives a case of lightness number being 0.85. The problem is optimized for minimum solar distance between 0.25 AU and 0.5 AU. The impact decreases shapely as the minimum distance increases while the transfer time almost keeps fixed. Therefore, as long as the sail material can bear the environment, the solar sail should approach the Sun as close as possible. Figure 4 gives a case of the minimum solar distance being 0.3 AU. The problem is optimized for lightness number between 0.75 and 0.9. The impact velocity increases with the lightness number linearly while transfer time increases slowly with the lightness number. It can be concluded from the simulations that a small minimum solar radius and a large lightness number lead to high impact velocity. The transfer time changes with slowly with the minimum solar radius and lightness number.

Figure 5 shows a typical impact trajectory and figure 6 gives the corresponding parameters along the trajectory, where $\eta$ is the angle between the velocity of the sail and normal of the sail. The sail accelerates for $\eta < 90^\circ$ and decelerates for $\eta > 90^\circ$. The time for acceleration and deceleration are similar during the journey. Before the point where $\eta = 90^\circ$, the variation of energy of the sail is small. However, the energy increases quickly when the sail approaches the Sun, which can be seen from the energy history of the sail. That's because the acceleration ability increases as the sail approaches the Sun.

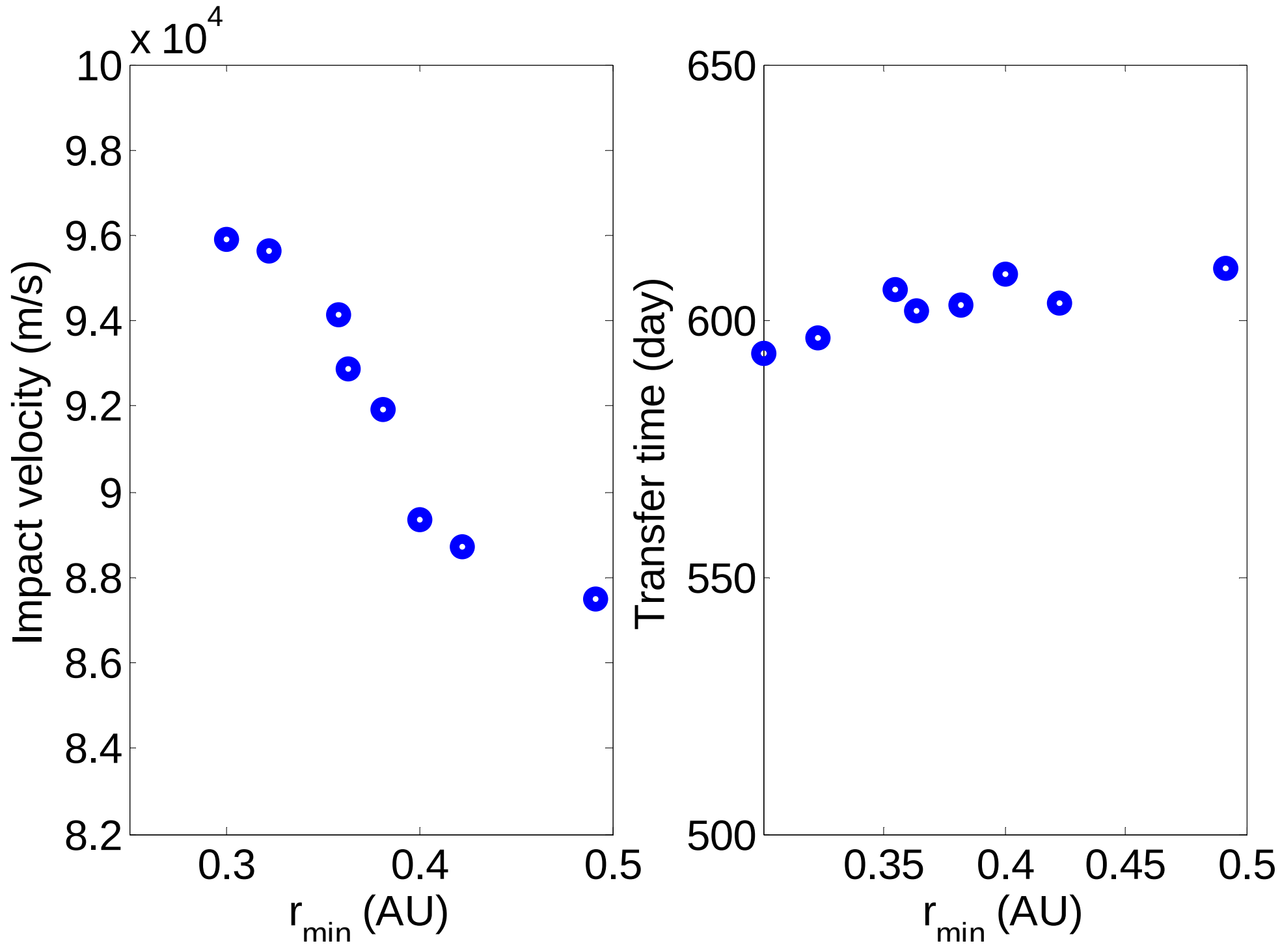

Fig.3 Impact velocity and transfer time for different minimum solar radius, $\beta = 0.85$

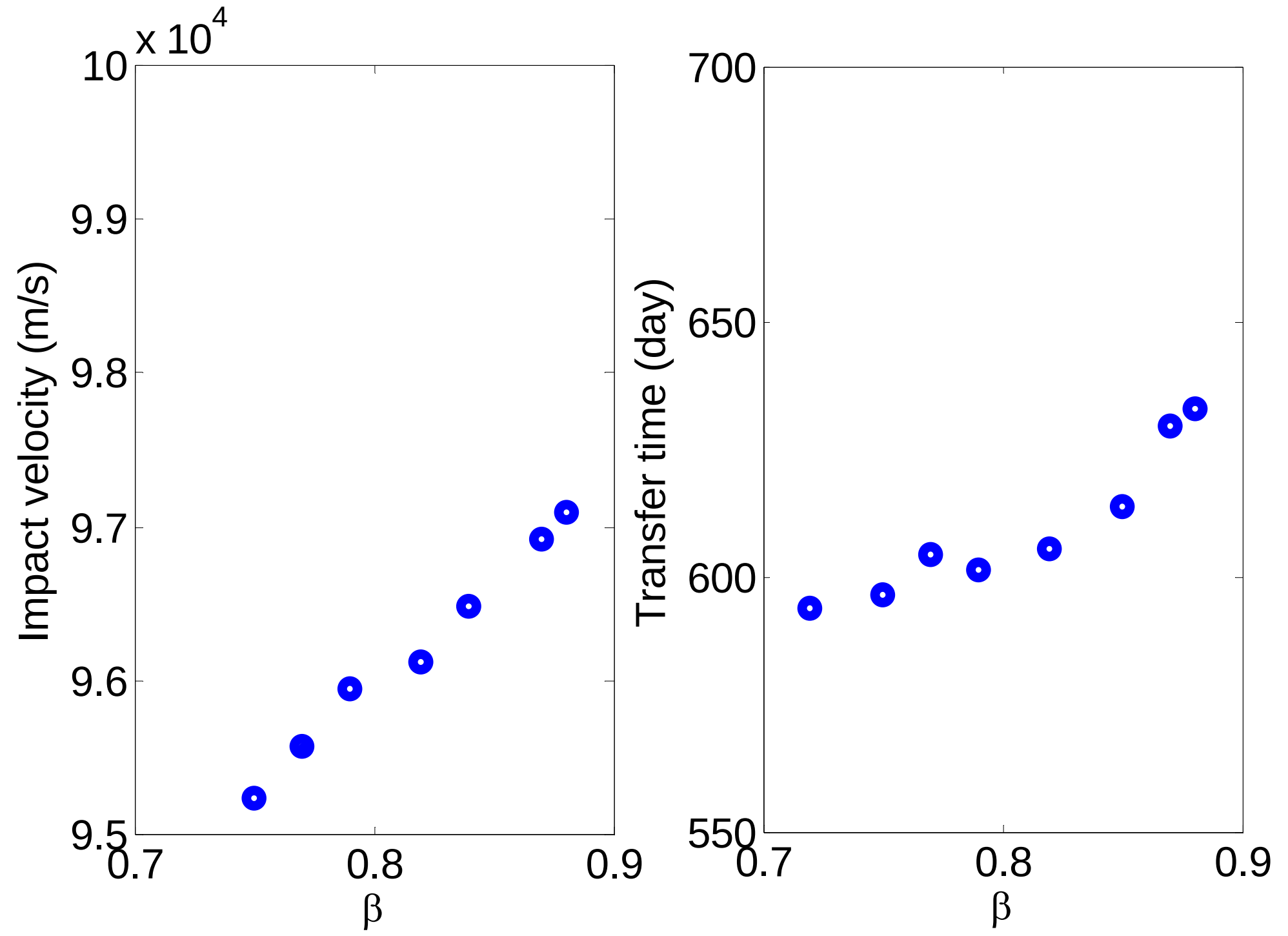

Fig.4 Impact velocity and transfer time for different lightness number, $r_{\min} = 0.3$ AU

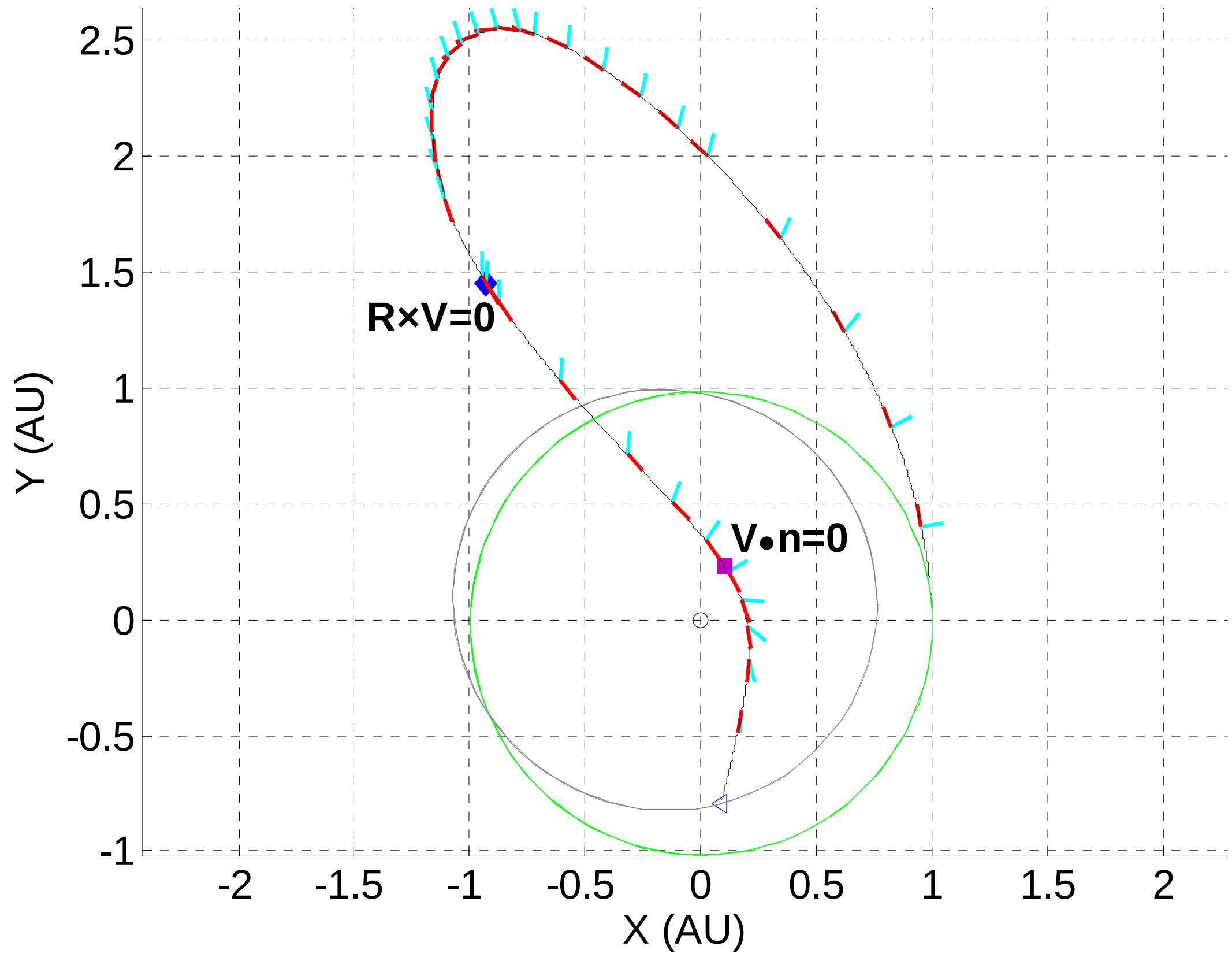

Fig.5 An impact trajectory with Apophis using H-reversal trajectory

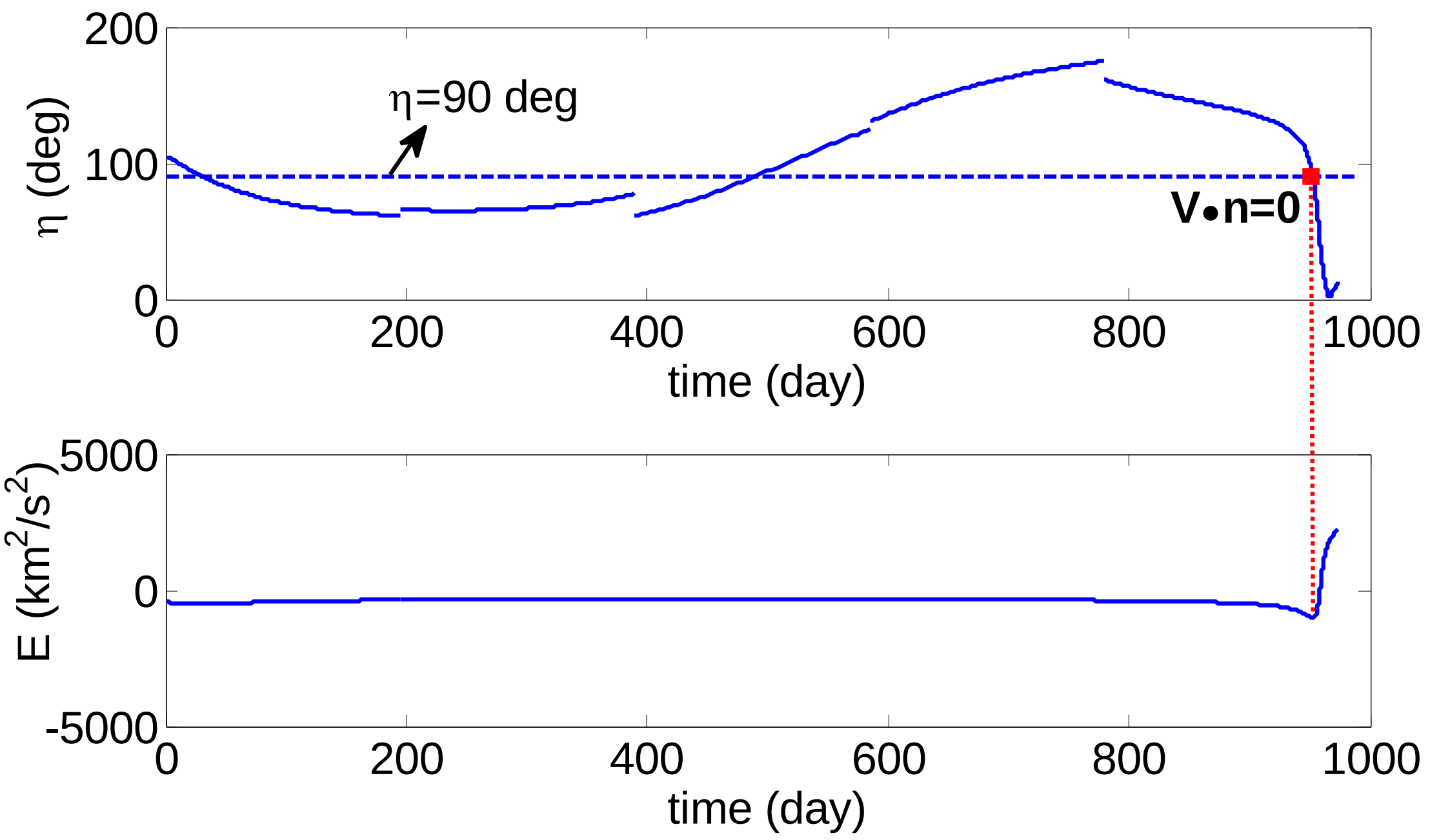

Fig.6 histories of mechanical energy and angle $\eta = \cos^{-1}\left(\frac{\boldsymbol{V} \cdot \boldsymbol{n}}{|\boldsymbol{V}|}\right)$

**5. Deflection Capability Discussion**

It is assumed that the momentum of the system is conserved during the impact. The impact is almost a head to head impact. A scalar equation of conservation of momentum is used here to describe the velocity change of the asteroid along the velocity direction.

$$MV_1 + mV_2 = (M + m)V_3 \tag{9}$$

where $M$ and $m$ are the mass of the asteroid and solar sail, respectively; $V_1$ and $V_2$ are the velocity of the asteroid and solar sail before impact, respectively; $V_3$ is the velocity of the asteroid after impact.

Then, the change in speed to be delivered to the asteroid can be obtained as

$$\Delta V = V_1 - V_3 = \frac{m}{M}(V_3 - V_2) \tag{10}$$

The velocity of the asteroid before and after impact is very close. It means that $V_3 - V_2$ is very close to the impact velocity of the solar sail. Therefore, the change in the speed of the asteroid is determined by the mass ratio and impact velocity.

The asteroid deflection is determined by the lead-time and the speed change of the asteroid. The formula used to calculate the asteroid deflection can be given as (Wie, 2007)

$$\Delta L = \Delta V \cdot \Delta t \tag{11}$$

where $\Delta t$ is the lead-time.

The utility of H-reversal trajectory can raise the impact velocity to about 100km/s. For an impact velocity of 90 km/s, the deflection capability for different sail mass and lead-time is shown in Fig.7. A 140kg solar sail with a lead-time of 20 year generates a deflection of about 140km. To move Apophis out of a 600-m keyhole area in 2029 to eliminate the possibility of its resonant return in 2036 requires only a 10 kg solar sail with a lead-time of one year. Compared with a regular spacecraft, solar sail using H-reversal trajectory requires less mass. The impact velocity of a typical spacecraft along a Kepler orbit is about 30 km/s. A solar sail evolving a retrograde Kepler orbit can raise the impact velocity to about 60 km/s. The solar sail utilizing a H-reversal trajectory raises the impact velocity further to about 90 km/s. Therefore, the impact energy can be greatly enhanced for unit mass and the impact efficiency is much higher. One week point of a H-reversal is that a high performance solar sail is required. It means that much larger area of sail film is required for the same mass of spacecraft. The solar sail leaves the Sun at a high velocity. To impact with the asteroid has a high demand on the navigation, guidance and control system since a small error may make the sail miss the asteroid.

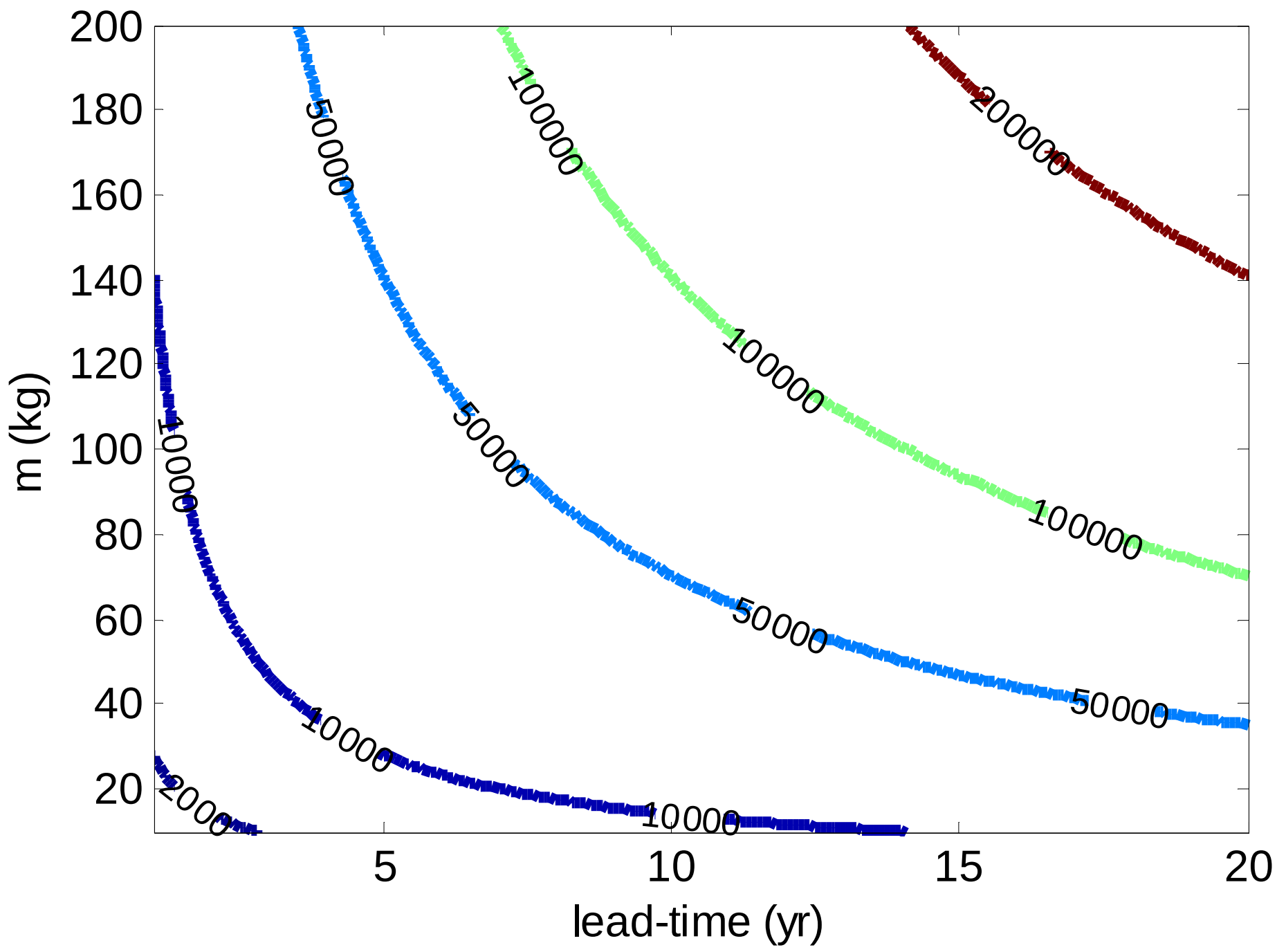

Fig.7 deflection capability of utilizing H-reversal trajectory of solar sail (unit, m)

## 6. Conclusion

A high performance solar sail can evolve in a H-reversal trajectory. A typical H-reversal trajectory is realized by reducing the angular momentum of solar sail until it is reversed. Then, the solar sail approaches the Sun to gain energy and leaves the Sun along a hyperbolic trajectory and impacts with the asteroid head-to-head. An optimization method is utilized to maximize the impact velocity. The impact velocity is dependent on the minimum solar distance along the trajectory and lightness number. For minimum distance less than 0.3 AU, the impact velocity is above 90 km/s. For this impact velocity, a solar sail of 10 kg with a lead-time of one year can move the Apophis out of its 600-m keyhole area.

**Acknowledgements** This work was funded by the National Natural Science Foundation of China(NSFC; Grant Nos. 10902056 and 10832004).

## References

Melosh, H.J. 1993, Nature, 366, 21.